\documentclass[%
 reprint,
 amsmath,amssymb,
 aps,
]{revtex4-1}

\usepackage{graphics}
\usepackage{graphicx}
\usepackage{dcolumn}
\usepackage{bm}

\usepackage[usenames,dvipsnames]{xcolor}

\usepackage{color}
\definecolor{darkgreen}{rgb}{0,0.6,0}
\definecolor{darkblue}{rgb}{0,0,0.6}
\definecolor{darkred}{rgb}{0.6,0,0}
\definecolor{darkpurple}{rgb}{0.5,0,0.5}

\usepackage{hyperref}

\hypersetup{
bookmarksopen=true,
pdftitle=From bulk to interfaces: connection between the Ginzburg--Landau and the elastic line models,
pdfauthor=, 
pdftoolbar=false,           
pdfstartview={FitH},		
pdfmenubar=true,			
pdfhighlight=/O,			
colorlinks=true,			
urlcolor=darkblue,
citecolor=darkblue,		    
linkcolor=darkpurple	    
}

\newcommand{\ff}{\text{f}}
\newcommand{\eff}{\text{eff}}

\newcommand{\argc}[1]{\left[#1\right]}
\newcommand{\arga}[1]{\left\lbrace #1\right\rbrace }
\newcommand{\argp}[1]{\left(#1\right)}
\newcommand{\valabs}[1]{\vert #1\vert}
\newcommand{\moy}[1]{\left\langle  #1 \right\rangle }

\def\de{\mathrm d}

\begin{document}

\preprint{APS/123-QED}

\title{From bulk descriptions to emergent interfaces: connecting the Ginzburg--Landau and elastic line models}

\author{Nirvana Caballero$^1$}
\email[Corresponding author: ]{Nirvana.Caballero@unige.ch}

\author{Elisabeth Agoritsas$^2$}

\author{Vivien Lecomte$^3$}

\author{Thierry Giamarchi$^1$}
\affiliation{$^1$Department of Quantum Matter Physics, University of Geneva, 24 Quai Ernest-Ansermet, CH-1211 Geneva, Switzerland}

\affiliation{$^2$Institute of Physics, Ecole Polytechnique Fédérale de Lausanne (EPFL), CH-1015 Lausanne, Switzerland}

\affiliation{$^3$Université Grenoble Alpes, CNRS, LIPhy, 38000 Grenoble, France}
\date{\today}
\begin{abstract}
Controlling interfaces is highly relevant from a technological point of view. However, their rich and complex behavior makes them very difficult to describe theoretically, and hence to predict. In this work, we establish a procedure to connect two levels of descriptions of interfaces: for a bulk description, we consider a two-dimensional Ginzburg--Landau model evolving with a Langevin equation, and boundary conditions imposing the formation of a rectilinear domain wall. At this level of description no assumptions need to be done over the interface, but analytical calculations are almost impossible to handle. On a different level of description, we consider a one-dimensional elastic line model evolving according to the Edwards--Wilkinson equation, which only allows one to study continuous and univalued interfaces, but which was up to now one of the most successful tools to treat interfaces analytically. To establish the connection between the bulk description and the interface description, we propose a simple method that applies both to clean and disordered systems. We probe the connection by numerical simulations at both levels, and our simulations, in addition to making contact with experiments, allow us to test and provide insight to develop new analytical approaches to treat interfaces.
\end{abstract}

\maketitle

\section{Introduction}
\label{sec:Intro}

Diverse systems including ferroic domain walls~\cite{cherifi2017non,ziegler2013domain,paruch2013nanoscale,guyonnet2012multiscaling,torres2019universal,lee2009roughness,caballero2017excess,magni2009visualization,ferre2013universal,grassi2018intermittent,Jordan2020},
cell fronts~\cite{huergo2010morphology,muzzio2014influence}, bacterial colonies~\cite{Bonachela_JSP_2011}, or contact lines~\cite{moulinet2004width} exhibit emergent structures separating different ``states'' or domains (\textit{i.e.},~different magnetization orientations in the case of ferromagnetic systems, or different polarization orientations in the case of ferroelectrics, or cells-media in cell fronts, or wet from dry in the case of contact lines), usually called interfaces. From a technological point of view, controlling interfaces is of great interest for various reasons. In some cases, interfaces are used as the base unit of devices (for example, in data storage devices~\cite{parkin2008magnetic}), and in others, interfaces are used to extract information about the whole system by simply observing a fraction of the system (for example in the case of cells colonies, where the interface gives information about the interactions present in the tissue~\cite{caballerocells2020}).

Interfaces have been usually described as disordered elastic systems (DES) \cite{giamarchi_2006_arXiv:0503437,agoritsas_review}. In this framework, interfaces are approximated by univalued and continuous functions of position and time.
In a great number of cases this is a good approximation since usually the region where the system changes from a state to another is small compared to the regions where the system is homogeneous. In particular, in the aforementioned systems, interfaces can be treated as unidimensional elastic objects, leading to a very simplistic description, which still captures the essential ingredients describing the physics of these objects. 

The advantage of treating interfaces as one-dimensional univalued functions is that it allows one to compute analytically, and in a very precise way, several observables and critical exponents describing dynamic and static properties of interfaces, allowing for a better understanding of their properties, and thus a better control over them.   
However, it is well known that real experimental realizations of interfaces are usually far away from being described by univalued functions, and in order to use the DES theoretical framework, uncontrolled approximations are used to force the real interface to be adapted to one of the main hypotheses of this framework.

On a different level of treatment for interfaces, Ginzburg--Landau (GL) models, where the state of the system is described by a local order parameter which can take real values in a well-defined range, can also describe interfaces, and the advantage is that assumptions about the function describing the interface are no longer needed. Moreover, effects like nucleation, bubbles, and non-univalued interfaces may arise, allowing for a more realistic description of interfaces. The lack of intrinsic periodic pinning, usually present in spin-like models, makes this approach extremely suitable for the study of interfaces. However, analytical calculations are very difficult to tackle for these kind of models.

Both levels of description, the elastic line model, and GL models have been proven helpful to describe the physics of disordered systems very well. However, a complete connection between the two levels of description, or `model reduction', is still lacking. Establishing a connection between both models is extremely important, since it allows one to obtain analytical predictions for the more complex model, based on results for its simpler counterpart.
This question is quite generic since the dynamics is that of the so-called `model A'~\cite{RevModPhys.49.435_hohehalp}.
A model reduction has been determined for flat walls in the absence of noise~\cite{allen_microscopic_1979},
or using a Fokker--Planck viewpoint~\cite{kawasaki_dynamics_1977,kawasaki_kink_1982}
or other approaches for
flat interfaces~\cite{PhysRevLett.47.1837_bausch,PhysRevB.28.5496_grant},
and in the context of 
kinetic roughening~\cite{PhysRevA.43.1727_grossmann}
or of the 
`drumhead model'~\cite{10.1143PTP.67.147_kawasaki,diehl_interface_1980}.
More complex approaches than the ones we propose have also been developed, including for instance the effect of curvature~\cite{PhysRevB.28.5496_grant,Zia1985_normalcoord,PhysRevA.43.1727_grossmann,PhysRevE.64.021604_kosterlitz} or of varying domain-wall width~\cite{10.1143PTP.67.147_kawasaki}.
Note that the model reduction is formally equivalent to the determination of extended particle states in quantum field theory~\cite{PhysRevD.12.1038_gervais,callan_quantum_1975}, 
where collective coordinate methods are similar to those of statistical mechanics.

In this work, we connect these two models through a simple procedure that requires few assumptions, and that applies both to clean systems and to systems with quenched disorder. 
This is a first step to get insight in how to extend the DES theory beyond the elastic approximation, thus allowing for a better characterization and understanding of experimental realizations of interfaces.
The plan of the paper is as follows. 
In Sec.~\ref{sec:FromBulktoInterface}, we briefly describe the GL model, establish the necessary assumptions and propose a procedure to connect this model to an Edwards--Wilkinson (EW) elastic line model, in the clean case. Complementary justifications of our procedure are presented in Appendices~\ref{sec:Tstar}
to~\ref{sec:PI-nodis}.
In Sec.~\ref{sec:Roughness} we compute analytically how the roughness, an observable measuring geometrical fluctuations of an interface, evolves as a function of lengthscale and time for a 1D elastic line. We probe the established connection between the models by performing simulations on a 2D-GL model, a 1D-EW model: we evaluate the roughness of interfaces which evolved starting from a completely flat configuration, and show how interfaces in both models, under our proposed connection, behave in excellent agreement with the analytical prediction in the 1D case. We also probe the connection between models numerically as a function of temperature. In Sec.~\ref{sec:Disorder}, we introduce quenched disorder in the GL system and show how it translates in the EW model into a short-range correlated disorder. We evaluate numerically the roughness and its Fourier transform, the structure factor, and show that they are in excellent agreement in both models, validating our proposed procedure for disordered systems. We finally conclude and discuss some perspectives of our work in Sec.~\ref{sec:Conclusions}.

\section{From bulk dynamics to interface dynamics (clean systems)}
\label{sec:FromBulktoInterface}

\begin{figure}
\begin{center}
{\includegraphics[width=1\linewidth]{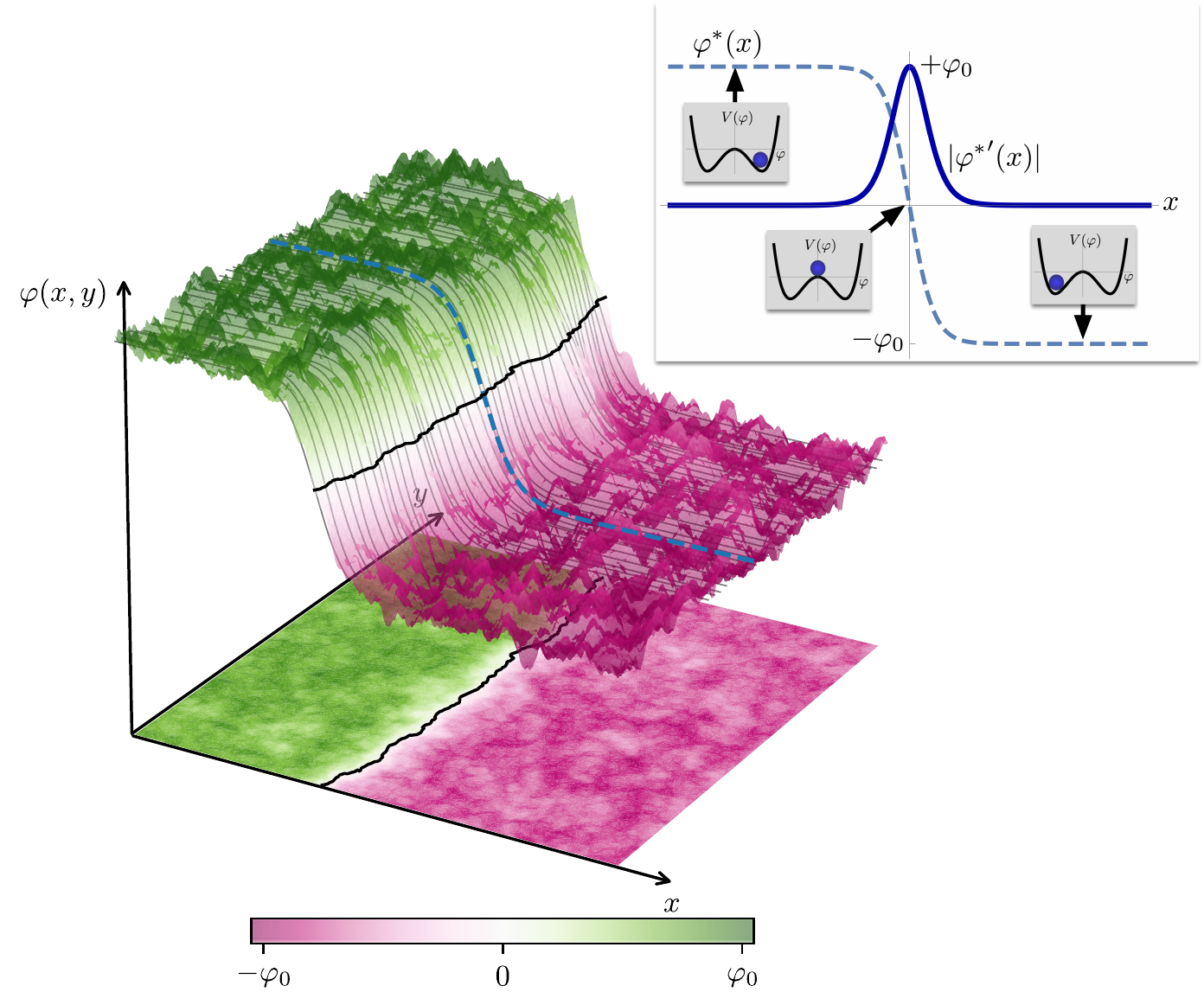}}
\end{center}
\caption{
    Snapshot of part of a system after solving numerically the Langevin equation (see text) for a 2D Ginzburg--Landau model (Eq.~\eqref{eq:Langevin}, with $\eta=\alpha=\delta=\gamma=1$, $T=0.05$, $t=10^5$) to obtain the evolution of the order parameter $\varphi(x,y)$. The obtained interface for this system is also shown in black. One of the fitted soliton profiles $\varphi^*(x)$ (for fixed $y$) is highlighted in dashed blue line. On the inset: the hyperbolic profile $\varphi^*(x)$ from Eq.~\eqref{eq:Soliton}, its derivative (which characterizes the 
   `density' of the interface), and three typical states in the local double-well potential.}
  \label{fig:wall3d}
\end{figure}

We study the behavior of the region (or `interface') separating  two domains characterized by distinct values of the local order parameter in a bulk model (see Fig.~\ref{fig:wall3d}). At the bulk level, we use a Ginzburg--Landau (GL) model to describe the system, where the order parameter of each homogeneous region is a local minimum of the corresponding ``$\varphi^4$'' potential. We consider a non-conserved order parameter, $\varphi(\mathbf{r},t)$, describing the local state of the system ruled by a GL Hamiltonian
\begin{equation}
\begin{aligned}
 \mathcal{H}_{\text{GL}}[\varphi]&= \int \de {\mathbf{r}} \, \argc{\frac{\gamma}{2}|\nabla_{{\mathbf{r}}} \varphi|^2 + V(\varphi) -h\varphi }, \\
\label{eq:Hamiltonian}
\end{aligned}
\end{equation}
where ${\mathbf{r}}\in \mathbb{R}^n$, and the $\varphi^4$ potential
\begin{equation}
V(\varphi)= -\frac{\alpha}{2}\varphi^2+\frac{\delta}{4}\varphi^4
\label{eq:V}
\end{equation}
with $\alpha>0$, $\delta>0$,
models the existence of two preferred values for $\varphi$: the minima of this double-well potential at ${\pm \varphi_0 =\pm \sqrt{\alpha/\delta}}$ represent the two preferential states of the system, and $h$ is an external applied field.

In this section, to establish the procedure, we focus on a clean system. The effect of disorder, which is crucial for experimental realization of interfaces, will be studied in details in Sec.~\ref{sec:Disorder}. 

The simplest equation describing the time evolution of the non-conserved order parameter $\varphi(\mathbf{r},t)$ in contact with a thermal bath at temperature $T$ is given by the overdamped Langevin equation
\begin{equation}
\eta \partial_t \varphi
=
-\frac{\delta \mathcal{H}_{\text{GL}}[\varphi]}{\delta \varphi}+\xi
= 
\gamma\nabla_{{\mathbf{r}}}^2\varphi -V'(\varphi)  + h +\xi \, , \\ 
\label{eq:Langevin}
\end{equation}
where $\xi=\xi(\mathbf{r},t)$ is a Gaussian white noise with zero mean and two-point correlator
\begin{equation}
\langle \xi({\mathbf{r}}_2,t_2)\xi({\mathbf{r}}_1,t_1) \rangle=2 \eta T \delta^n({\mathbf{r}}_2-{\mathbf{r}}_1)\delta(t_2-t_1),
\label{eq:thermalcorrelations}
\end{equation}
$\eta$ is the microscopic friction and ${\gamma}$ the amplitude of the elastic cost  associated to deformations of~$\varphi$.

Interfaces are defined as the region where the order parameter shifts from a preferred value to another. We are interested in studying interfaces in a 2D system with $\mathbf{r}=(x,y)$ (see Fig.~\ref{fig:wall3d}).
To do so, if the $x$ and $y$ axes are chosen so that the interface has a univalued shape at ${x=u(y,t)}$,
a natural ansatz to describe the field is $\varphi(x,y,t)=\varphi^*(x-u(y,t))$, where the function  $\varphi^*$ describes the switch from a preferred value of the order parameter to another.
Such an ansatz can only be approximate since, at non-zero temperature, the actual shape of the switching profile actually depends on the $y$ coordinate and presents fluctuations of thermal origin (see Fig.~\ref{fig:wall3d}).
We expect it to become correct at low temperature if the function $\varphi^*$ is well chosen.
As shown in Appendix~\ref{sec:Tstar}, the thermal fluctuations of the order parameter $\varphi(x,y,t)$ in each of the $\pm\varphi_0$ phases are negligible compared to their mean value if the temperature is much lower than $T^\star=\alpha\gamma/\delta$. 
We thus expect our analysis to be valid in the regime $T\ll T^\star$ (see Ref.~\cite{bausch_effects_1991} for a treatment of thermal fluctuations in the bulk).
In order to determine an effective equation of evolution for the so-called displacement field ${u(y,t)}$,
we substitute the ansatz into the bulk Langevin Eq.~\eqref{eq:Langevin}:
\begin{align}
-\eta {\varphi^{*}}'\partial_t u
 &=
 \gamma\Big( {\varphi^{*}}''+ {\varphi^{*}}''(\partial_y u)^2-{\varphi^{*}}'\partial^2_y u\Big)
 \nonumber
 \\
 & \quad
 - V'(\varphi^*)+h+\xi.
\label{eq:Langevinansatz1}
\end{align}

Physically, we expect that at low temperature the optimal $\varphi^*$  is a solitonic profile that minimizes the energy of the system at zero field $h$:
\begin{equation}
-\frac{\delta \mathcal{H}_{\text{GL}}[\varphi]}{\delta \varphi}\Big|_{\varphi^*}
=
\gamma {\varphi^{*}}''-V'(\varphi^*)
= 
0.
\label{eq:Minimization}
\end{equation}
Such an equation effectively describes the conservative motion of a ``particle'' of position $\varphi^*$ and time $x$ that evolves in a potential $V$.
If the function $V(\varphi)$ has two local minima, we indeed have solitonic solutions that go from a minimum to another as $x$ goes from $-\infty$ to $+\infty$.
In our case of interest~\eqref{eq:V}, we pick the soliton that satisfies the Dirichlet boundary conditions $\varphi^*(\pm \infty)=\mp\varphi_0$ whose explicit form is well known:
\begin{equation}
 \varphi^*(x)=-\varphi_0 \tanh \Big(\frac{x}{w}\Big),
 \label{eq:Soliton}
\end{equation}
as illustrated in Fig.~\ref{fig:wall3d}.
The parameters $w$, representing the width of the interface, and $\varphi_0$, representing the preferred values $\pm\varphi_0$ for the order parameter are given by
\begin{equation}
\varphi_0=\sqrt{\frac{\alpha}{\delta}} \,,\,\,\,\,\,\,\, w=\sqrt{\frac{2\gamma}{\alpha}}.
\label{eq:SolitonParameters}
\end{equation}

Substituting the identity~\eqref{eq:Minimization} into Eq.~\eqref{eq:Langevinansatz1},
one obtains explicitly
\begin{align}
-\eta {\varphi^{*}}'(x)
&\partial_t u(y,t)
\nonumber
\\
&=
 \gamma \argc{ {\varphi^{*}}''(x)\big(\partial_y u(y,t)\big)^2-{\varphi^{*}}'(x)\partial^2_y u(y,t)}
\nonumber
\\
&
\quad
+h+\xi(x+u(y,t),y,t),
\label{eq:Langevinansatz1bis}
\end{align}
where we can safely replace $\xi(x+u(y,t),y,t)$ by $\xi(x,y,t)$ using the invariance by translation of the noise distribution.

The equation of evolution~\eqref{eq:Langevinansatz1bis} is inconsistent (the dependency in $x$ is not the same for every term), even at zero temperature.
To obtain an equation of evolution for the position of the interface, one multiplies Eq.~\eqref{eq:Langevinansatz1bis} by ${\varphi^{*}}'$,
in order to ``localize'' the equation around the position of the interface, and one integrates 
over $x$. A justification of this procedure is presented in Appendix~\ref{sec:GL-DES-Hamiltonian-solitonic-ansatz} (see Eq.~\eqref{eq-chain-rule-functional-DEs-force}):
at the energetic level, when computing the force as deriving from a bulk or an effective Hamiltonian, a factor ${\varphi^*}'$ naturally appears between the derivatives $\frac{\delta}{\delta u}$ or $\frac{\delta}{\delta\varphi^*_u}$.
See also Appendix~\ref{sec:PI-nodis} for a path-integral approach where the integration over $x$ comes naturally, directly in a dynamical formulation.
Doing so, one obtains
\begin{equation}
\eta \mathcal{N}_1\partial_t u
=
\gamma\mathcal{N}_1\partial^2_y u - \gamma\mathcal{N}_2 (\partial_y u)^2 + h\mathcal{N}_3 +\tilde \xi(y,t), 
\label{eq:langevinansatz2bis}
\end{equation}
where
\begin{equation}
 \mathcal{N}_1\equiv \int_{-\infty}^{\infty}  \!\! \de x \, ({\varphi^{*}}')^2= \varphi^2_0 \frac{4}{3w}=\frac{2\sqrt{2}}{3\delta}\sqrt{\frac{\alpha^3}{\gamma}},
\label{eq:defN1-DW}
\end{equation}
\begin{equation}
 \mathcal{N}_2\equiv \int_{-\infty}^{\infty}  \!\! \de x \,  {\varphi^{*}}''{\varphi^{*}}' =0, \quad \mathcal{N}_3=\int_{-\infty}^{\infty} \!\! \de x \, {\varphi^{*}}'=-2\varphi_0.
\label{eq:defN2N3-DW}
\end{equation}
The effective noise
\begin{equation}
 \tilde\xi(y,t)= \int_{-\infty}^{\infty}  \!\! \de x \, \xi(x,y,t) {\varphi^{*}}'(x)
\end{equation}
is a linear superposition of Gaussian variables, and is thus also a Gaussian white noise of zero average and 
correlations
\begin{equation}
 \langle \tilde \xi(y_2,t_2) \tilde\xi(y_1,t_1)\rangle
=  2 \eta T \mathcal{N}_1\delta(y_2-y_1) \delta(t_2-t_1).
\end{equation}
We thus find a Langevin equation for ${u(y,t)}$ of the form
\begin{equation}
\tilde \eta \partial_t u= c\partial^2_y u+ F +\tilde \xi, 
\label{eq:EW}
\end{equation}
which is the EW equation~\cite{edwards_wilkinson} describing the time evolution of an elastic line $u(y,t)$, with friction $\tilde \eta$, elasticity $c$, external force $F$, and temperature~$T$. By this procedure, we found the friction and the force effectively ``felt'' by an interface in the GL model, as well as its elastic constant, and how these quantities are related with the model parameters as
\begin{equation}
\begin{aligned}
\tilde \eta&\equiv \eta \mathcal{N}_1=\eta\frac{2\sqrt{2}}{3}\frac{\alpha}{\delta}\sqrt{\frac{\alpha}{\gamma}}
\, , \\ 
c&\equiv \gamma \mathcal{N}_1=\frac{2\sqrt{2}}{3}\frac{\alpha}{\delta}\sqrt{\alpha\gamma}
\, , 
\\
F&\equiv h\mathcal{N}_3=-2\sqrt{\frac{\alpha}{\delta}}h.
\label{eq:ParametersAnalogies}
\end{aligned}
\end{equation}
Note that the sign of the drive $F$ does depend on the explicit choice of soliton in Eq.~\eqref{eq:Soliton}:
this is expected because the GL field $h$ favors the $+\varphi_0$ phase and will act with opposite sign on the other possible soliton $+\varphi_0 \tanh (x/w)$.
On the other hand, $\tilde\eta$ and $c$ are always defined as positive, and their numerical prefactors depend on the specific normalized density of the interface ${\rho_w(x) \propto \valabs{\varphi{^*}'(x)}}$ (see Appendix~\ref{appendix-normalized-density}).

By using the solitonic profile $\varphi^*$ (Eq.~\eqref{eq:Soliton}) as an ansatz to solve the Langevin equation for the GL model, we found a procedure to go from the two-dimensional description of the problem to an effective one-dimensional one. 
Interestingly the same relation between the elasticity $c$ of a domain wall in a one-dimensional system and the GL parameters can be obtained by computing the energy cost $E_{\text{el}}$ of the creation of a domain wall in the system, as was obtained before (see e.g.~\cite{allen_microscopic_1979}). 

\begin{figure*}[!p]
\begin{center}
 {\includegraphics[width=1\textwidth]{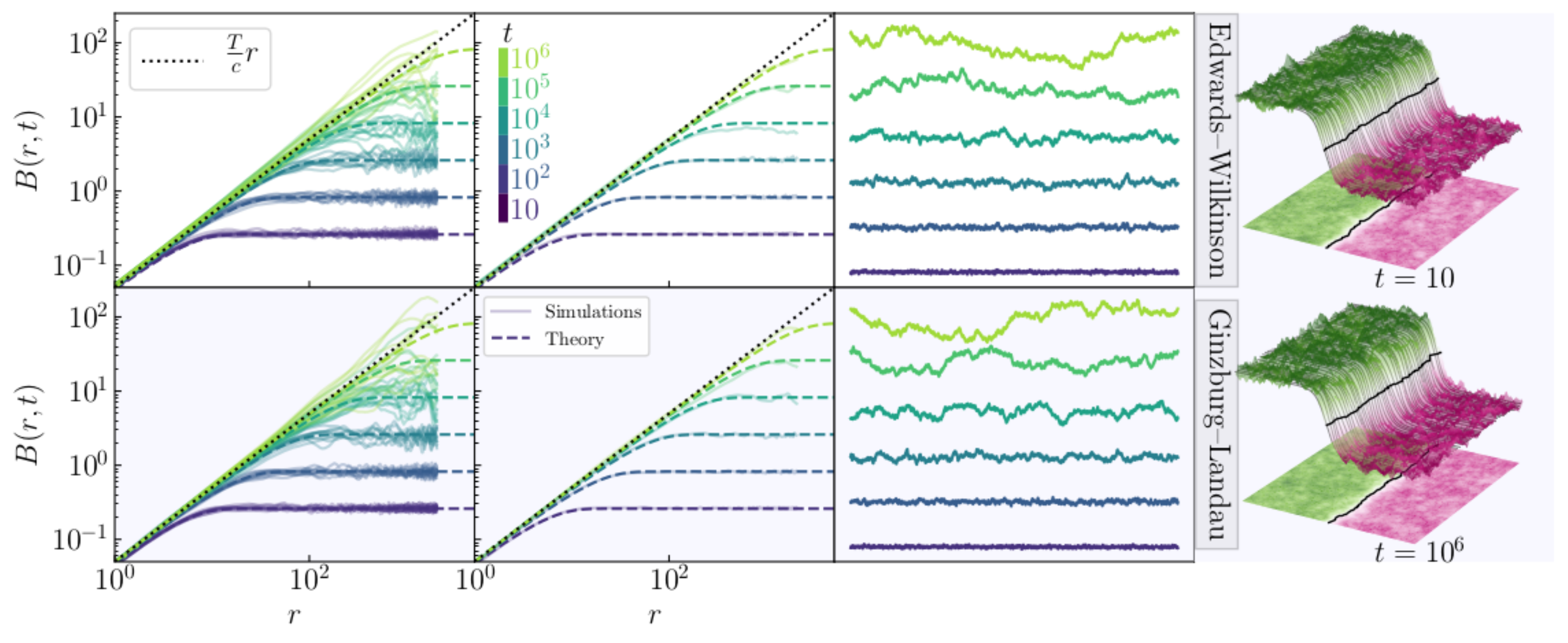}}
\end{center}
\caption{Time dependence of the roughness ${B(r,t)}$, computed for interfaces in a 2D Ginzburg--Landau system (bottom figures) and for an equivalent 1D Edwards--Wilkinson system (top figures), obtained for 10 realizations (left figures) and for the average over 10 realizations of simulations which evolved during a time $t=10^j$, $j=1, \cdots, 6$ (indicated by different colors), starting from a completely flat configuration. The analytical prediction of the evolution of $B(r,t)$ (Eq.~\eqref{eq:B(r,t)fromFlat}) for an equivalent one-dimensional interface is shown on dashed colored lines for different evolution times. The asymptotic value $\frac{T}{c}r$, expected for a completely stationarized interface, is shown in black dotted lines. On the right, the final extracted interfaces for one of the realizations after each evolution time $t$ are shown for both models. A portion of length $25.6\times25.6$ of a Ginzburg--Landau simulated system is also shown after evolution times of $t=10$ and $t=10^6$, along with the detected interface.}
  \label{fig:Br}
\end{figure*}

In this section we showed how to connect the GL and the DES descriptions at the level of their respective Langevin equation.
Our approach complements the one proposed in Ref.~\cite{PhysRevB.28.2693_safran} where both the elasticity and the thermal noise are also taken into account, but with a much more phenomenological treatment of the effective thermal noise.
The method we propose provides us with an effective reduced dynamics for the interface displacement field ${u(y,t)}$, that we test numerically in the subsequent sections, on the evolution of roughness starting from a flat initial condition. We will discuss this procedure in presence of disorder in Sec.~\ref{sec:Disorder}.

In Appendix~\ref{sec:GL-DES-Hamiltonian-solitonic-ansatz} we present a generic discussion on the model reduction from an equilibrium Hamiltonian viewpoint that complements the dynamical approach presented in this section. We show that the connection between the GL and the DES descriptions can actually be performed directly at the level of the Hamiltonian as well, if the system is assumed to be at equilibrium.
This is thus relevant for the long-time limit of the equilibrium dynamics (\textit{i.e.}~Eq.~\eqref{eq:Langevin} with no external field ${h=0}$), for which the probability of a given profile ${\varphi}$ is simply given by a Gibbs--Boltzmann distribution.
This procedure on the \emph{statics} allows us to identify the DES elastic constant $c$ and the effective disorder, but it does not give us access to the effective DES friction and noise since those pertain to the \emph{dynamics}, so we need to consider the Langevin equation as we did in this section (see also Appendix~\ref{sec:PI-nodis}).
Note also that the passage from Eq.~\eqref{eq:Langevinansatz1bis}
to Eq.~\eqref{eq:langevinansatz2bis}
bears similarity with the projection operator of Refs.~\cite{10.1143PTP.67.147_kawasaki,PhysRevB.28.5496_grant,PhysRevA.43.1727_grossmann,PhysRevB.29.6266_grant-gunton}.

\section{Roughness of interfaces}
\label{sec:Roughness}

\begin{figure*}[!p]
\begin{center}
 {\includegraphics[width=1\textwidth]{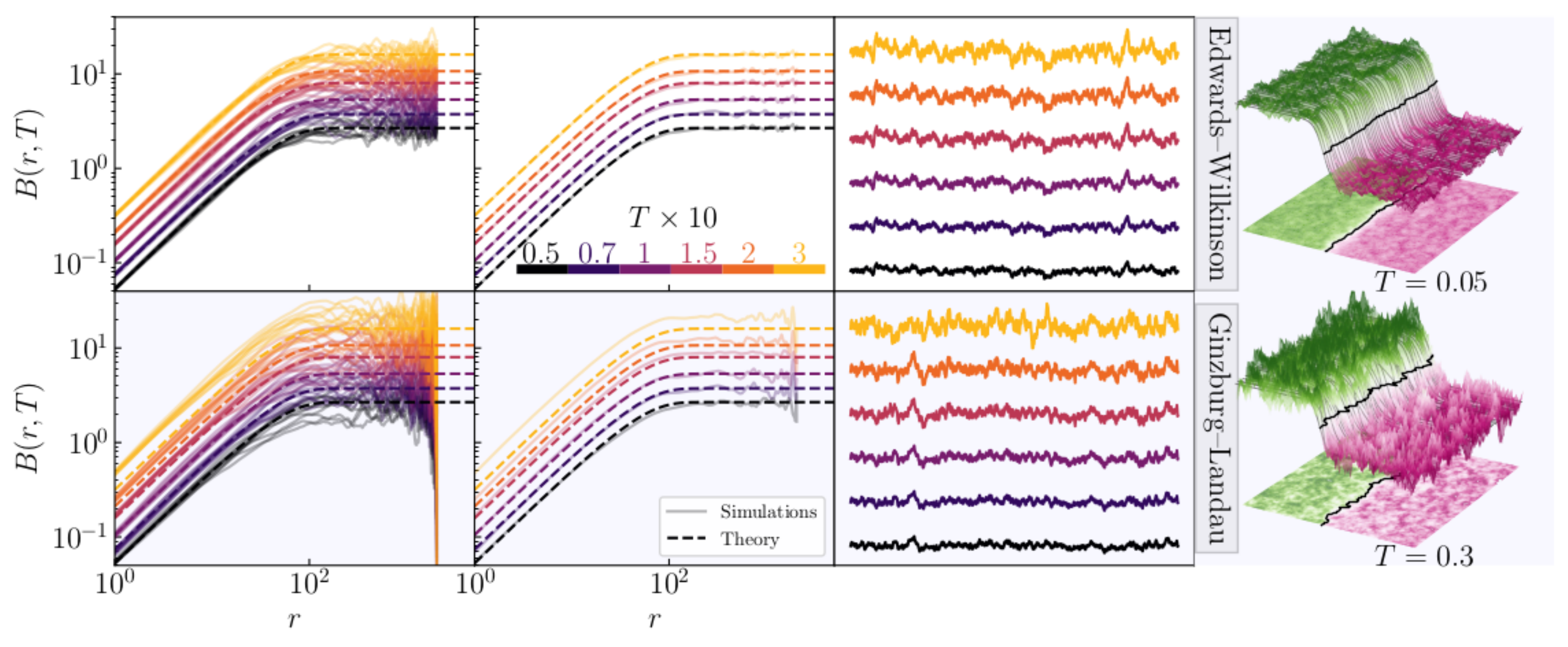}}
\end{center}
\caption{Temperature dependence of the roughness $B(r,T)$ for a 2D Ginzburg--Landau system (bottom figures) and for an equivalent 1D Edwards--Wilkinson system (top figures), obtained for 10 realizations (left figures) and for the average over 10 realizations of simulations which evolved during a time $t=10^3$ at temperatures $T=0.05,0.07,0.1,0.15,0.2,0.3$ (indicated by different colors), starting from a completely flat configuration. The analytical predictions of the evolution in time of $B(r,T)$ (Eq.~\eqref{eq:B(r,t)fromFlat}) for an equivalent one-dimensional interface is shown on dashed colored lines for different temperatures. The final interfaces obtained for one realization are also shown for both models at different temperatures. A portion of the Ginzburg--Landau system is also shown at $T=0.05$ and $T=0.3$, along with the detected interface, shown in black.}
 \label{fig:Br_temp}
\end{figure*}

Among the observables that characterize interfaces, one of the most useful, convenient, and studied is the one that measures the spatial correlations of the position  $u(y,t)$ of the interface at time $t$,
\begin{equation}
 B(r=|y_2-y_1|,t)= \overline{\langle [u(y_2,t)-u(y_1,t)]^2\rangle}.
 \label{eq:B(r)}
\end{equation}
This so-called roughness function characterizes the random geometry of the interface. $\langle \vphantom{|}\cdots \rangle$ denotes thermal average, and $\overline{\vphantom{|}\cdots}$ denotes the average over different disorder realizations when appropriate. Usually, it is also convenient to compute the Fourier transform, called structure factor, defined as 
\begin{equation}
 S(q,t)= \overline{
 \Big\langle \frac{1}{L} u_q^*(t) u_q(t)\Big\rangle}
 \, ,
 \label{eq:S(q)}
\end{equation}
where $u_q(t)=\sum^{L-1}_{j=0} (u_j(t)-\bar{u}(t))e^{iqj}$ (${\bar{u}(t)}$ is the mean position of the whole interface, zero thereafter),  and the discrete Fourier modes ${q=2\pi n/L}$ with ${n=1,\dots,L-1}$.

When a flat domain wall is subjected to a thermal bath, correlations in its geometry  evolve in time as a result of the competition between the domain wall elasticity and the thermal fluctuations. For finite times, a memory of the initial condition remains in Eq.~\eqref{eq:B(r)}. As $t$ goes to infinity, if the interface has a finite length, correlations spread along the whole interface, and this memory of the initial condition disappears.

For the clean system we are considering so far, we can compute analytically the full time dependence of this correlation. One uses the linearity of the EW equation to solve Eq.~\eqref{eq:EW} for ${F=0}$~\cite{edwards_wilkinson}, with an initially flat configuration. Averaging over the thermal noise, one obtains:
\begin{equation}
\begin{aligned}
B(r,t)
&=\frac{Tr}{c} \bigg[1- \frac{1}{\sqrt{\pi}zr}\Big(e^{-z^2r^2}-1\Big)- \frac{2}{\sqrt{\pi}}\int_0^{zr}e^{-t^2}dt \bigg]
\label{eq:B(r,t)fromFlat}
\end{aligned}
\end{equation}       
where $z=\sqrt{\frac{\tilde \eta}{8ct}}$. At large times, Eq.~\eqref{eq:B(r,t)fromFlat} converges to the static thermal roughness ${B_{\text{th}}(r) \equiv Tr/c}$.

We now use the result of Eq.~\eqref{eq:B(r,t)fromFlat} to assess the validity of our bulk-to-line model reduction.
To compare the numerical efficiency of the 2D-GL and of the 1D-EW modelisations, we first perform simulations of the 1D interface, \textit{i.e.}~we solve numerically Eq.~\eqref{eq:EW}~\cite{Ferrero2013nonsteady} with parameters $\tilde \eta=c=\frac{2\sqrt{2}}{3}$ (taking $\eta=\alpha=\delta=\gamma=1$ in Eq.~\eqref{eq:ParametersAnalogies}), ${T=0.05}$, and ${F=0}$~\footnote{We consider a discrete mesh along the $y$-direction, with mesh-size ${\ell=1}$ for a system of length ${L=4096}$, while keeping a continuum value of $u$. We use the Euler method to integrate the equation with a time step $\Delta t=10^{-2}$.}. Starting from a flat configuration, we perform simulations of the elastic line during different times for different realizations. For each final configuration obtained for $u(y,t)$ we compute $B(r,t)$. In Fig.~\ref{fig:Br} we show the obtained roughness functions for each realization and for an average of $B(r,t)$ over different realizations. We find an excellent agreement between the numerically obtained roughness functions and the analytical result~\eqref{eq:B(r,t)fromFlat}.

The analytical prediction for the roughness function given by Eq.~\eqref{eq:B(r,t)fromFlat} gives us a benchmark to test the proposed connection between the GL model of Eq.~\eqref{eq:Langevin} and the EW dynamics of Eq.~\eqref{eq:EW} in two and one dimensions respectively.
We performed simulations of a 2D-GL system, by solving numerically Eq.~\eqref{eq:Langevin}, with $\alpha=\delta=\gamma=\eta=1$, at ${T=0.05}$, with periodic boundary conditions along $y$ (interface direction), and Dirichlet boundary conditions along $x$ (see Fig.~\ref{fig:wall3d})~\footnote{for a system of length $L_x=256$, $L_y=4096$, with spatial mesh-size $\ell=1$ (in both directions). We integrate the equation by following a semi-implicit Euler method with integration time-step $\Delta t=0.1$ in Fourier space~\cite{jagla2004,caballero2018magnetic}.}.

Let us define for convenience the bulk order parameter $\varphi_u(x,y) = \varphi^*(x-u(y))$ associated to an interface of position $u(y)$ and a solitonic profile given by Eq.~\eqref{eq:Soliton} at each~$y$.
In the simulation, we start with a flat domain wall, \textit{i.e.}~with an initial condition $\varphi(x,y,t=0)=\varphi_{u_0}(x,y)$, with $u_0(y)=L_x/2$, for all $y$.
The order parameter $\varphi(x,y,t)$ then evolves in time by keeping the shape of a rectilinear domain wall profile, localized along an interface of position  $u(y,t)$ (see Fig.~\ref{fig:wall3d}).

To obtain the effective interface position $u(y,t)$ for a given configuration $\varphi(x,y,t)$ of GL model, we fit $\varphi(x,y,t)$ at fixed $y$ and $t$ with a function $\varphi_u(x,y)$, with the fitting parameters $\lbrace \varphi_0,w,u(y)\rbrace$.
The interface position $u(y,t)$ is then given by the fitted value $u(y)$~\footnote{The fitting parameters $\lbrace\varphi_0,w\rbrace $ exhibit at low temperatures a Gaussian distribution (with mean $\mu$ and standard deviation $\sigma$) around the equilibrium values (Eq.~\eqref{eq:SolitonParameters}). At $T=0.05$ we find $\mu_{\varphi_0}=0.99$ and $\sigma_{\varphi_0}=0.01$ for $\varphi_0$, and $\mu_w=1.38$, $\sigma_w=0.25$ for $w$.
These means are in good agreement with the expected values $\varphi_0=1$ and $w=\sqrt{2}$ (see Eq.~\eqref{eq:SolitonParameters}).}.
A snapshot of part of a simulated system is shown in Fig.~\ref{fig:Br} along with the detected interface and some of the fitted interface positions $u(y,t)$. By following this method, we computed $u(y,t)$ for different realizations of simulations of a system which evolved for different times, and we computed the roughness defined on Eq.~\eqref{eq:B(r)} of these functions.

The obtained values of the roughness are shown in Fig.~\ref{fig:Br} for different realizations at each time, and also for the average of the roughness over different realizations. The roughness functions of the interfaces obtained in our simulations are in excellent agreement with the expected result after different evolution times.
For the pure system, this strongly supports that we have a very precise method to connect both levels of descriptions of interfaces, in the elastic approximation.

This mapping allows us to test for the deviations for the pure elastic description of the interface. For the 1D-EW model, where the elastic description is exact by construction, no deviation from the elastic description indeed occurs. This can be seen in Fig.~\ref{fig:Br_temp}, where we computed the roughness of interfaces which evolved during a time $t=10^3$ for different temperatures $T$ and compared it with the theoretical prediction~\eqref{eq:B(r,t)fromFlat} that we denote $B(r,T)$ to emphasize the temperature dependence. However, for the 2D-GL model, the measured roughness functions match the predicted roughness only when the ratio $T/T^\star$ is sufficiently small (see Appendix~\ref{sec:Tstar}), with $T^\star=\alpha\gamma/\delta=1$ for our parameter values. We observe deviations from the theoretically expected value of $B(r,T)$ for temperatures larger than ${T=0.15}$. Such discrepancy as temperature increases is expected, since the approach we proposed to go from the bulk to the line model is based on a small-noise hypothesis.

\section{Disordered systems}
\label{sec:Disorder}

\begin{figure}[ht!]
\begin{center}
 {\includegraphics[width=0.5\textwidth]{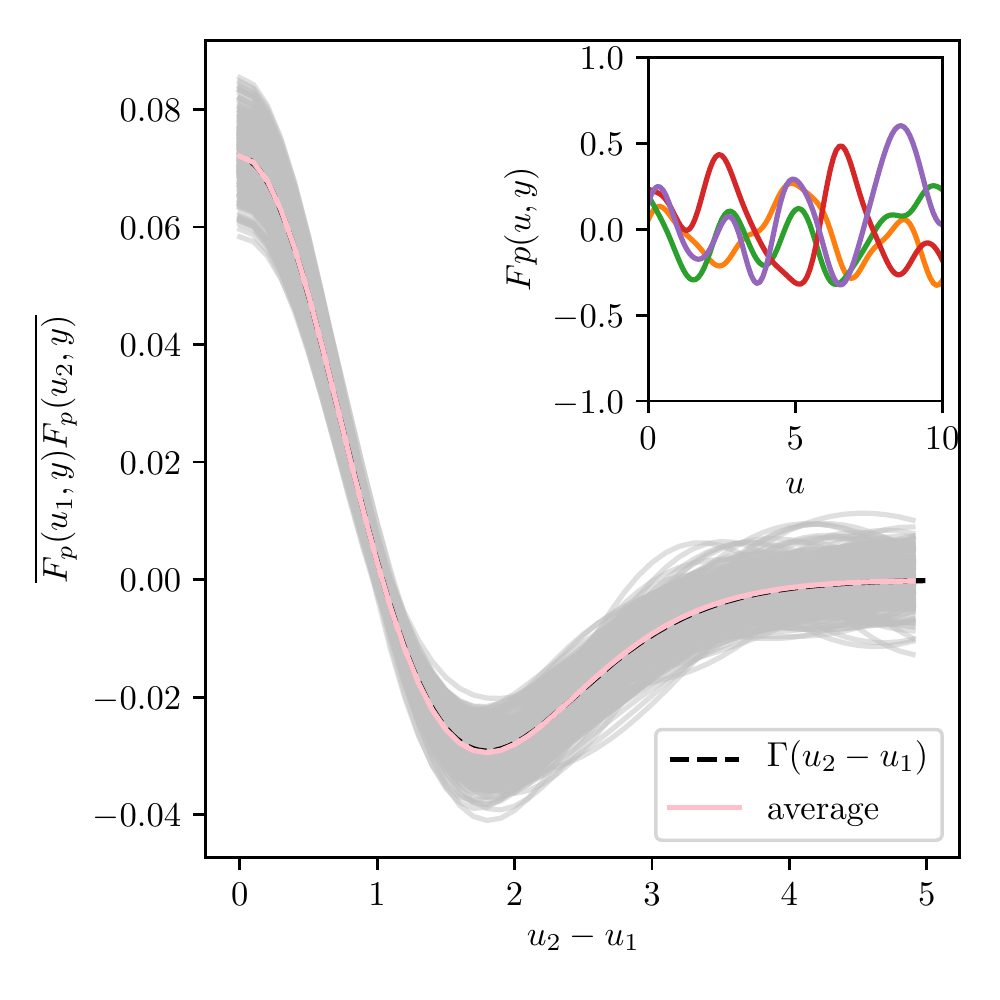}}
\end{center}
\caption{Computed correlations of 256 independent realizations of pinning forces $F_p(u,y)$ (in gray). The average of these correlations is plotted in pink, showing an excellent agreement with the expected correlations given by $\Gamma(u)$ (Eq.~\eqref{eq:N(u)}), shown in dashed black line. On the inset, 4 different realizations of pinning forces are shown.}
  \label{fig:Fp}
\end{figure}

Disorder plays a key role inducing highly non-linear effects in the statics and dynamics of interfaces. In particular, it is well known that, as a consequence of disorder, the interface geometry is drastically changed compared to one only subject to thermal fluctuations, and its study is the whole point of the DES framework \cite{ferre2013universal,giamarchi_2006_arXiv:0503437,agoritsas_review}. At small lengthscales, thermal fluctuations are expected to dominate the interface geometry behavior (at equilibrium $B(r) \approx B_{\text{th}}(r)=\frac{T}{c} r^{2\zeta_{th}}$, with $\zeta_{th}=1/2$). However, at large lengthscales, disorder induces a change in the power-law behavior of the roughness, and both, the prefactor and the roughness exponent $\zeta$ are affected~\cite{agoritsas_review,agoritsas2013static}. The equilibrium roughness $B(r)$ will thus be characterized at large distances by a different exponent dependent on the disorder type (for example random-bond or random-field types~\cite{chauve_creep_long}).
Let us now extend the mapping presented on Sec.~\ref{sec:FromBulktoInterface} to the case of disordered systems. 

To study the effect of quenched disorder on an interface described by a Ginzburg--Landau (GL) model, we introduce fluctuations in the height of the double-well potential $V(\varphi)$ of~\eqref{eq:V} as
\begin{equation}
V_\zeta(\varphi ({\mathbf{r}}))= V(\varphi({\mathbf{r}}))(1+\epsilon\zeta(\mathbf{r})).
\label{eq:Vdisorder}
\end{equation}
Here $\zeta(\mathbf{r})$ is a random number at position ${\mathbf{r}}$ taken from a Gaussian distribution with zero mean and unit variance, whose correlations satisfy  $\overline{ \zeta({\mathbf{r}}_i)  \zeta({\mathbf{r}}_j)}=\delta^2({\mathbf{r}}_i-{\mathbf{r}}_j)$, where ${\mathbf{r}}_{i,j}$ are the relative distance between the simulation cells $i$ and $j$, and we recall that $\overline{\vphantom{|}\cdots}$ denotes the average over different disorder realizations.

When using the ansatz $\varphi(x,y,t)=\varphi^*(x-u(y,t))$, the Langevin equation describing the evolution of the order parameter now becomes, instead of~\eqref{eq:Langevinansatz1bis},
\begin{align}
-\eta {\varphi^{*}}'\partial_t u
=&
\gamma\Big( {\varphi^{*}}''+ {\varphi^{*}}''(\partial_y u)^2-{\varphi^{*}}'\partial^2_y u\Big)
 \label{eq:langevinansatz1}
\\
&- V'(\varphi^*)-\epsilon\zeta(x,y)V'(\varphi^*) +\xi(x,y,t).
\nonumber
\end{align}
Following the procedure of Sec.~\ref{sec:FromBulktoInterface}, \textit{i.e.}~by multiplying by $-{\varphi^{*}}'$, using the soliton equation~\eqref{eq:Minimization} $\gamma {\varphi^{*}}''=V'(\varphi^*)$, and integrating $x$ over the whole space, we find an effective Langevin equation for the displacement field ${u(y,t)}$
\begin{equation}
\tilde \eta \partial_t u= c\partial^2_y u+ F_p(u(y,t),y)+F+\tilde\xi(y,t).
\label{eq:effu_disorder}
\end{equation}
Compared to Eq.~\eqref{eq:EW}, we have now the extra term
\begin{equation}
F_p(u,y)=\epsilon \gamma \int_{-\infty}^{\infty} \!\! \de x \, \zeta(x+u,y) {\varphi^{*}}''(x){\varphi^{*}}'(x),
\label{eq:Fp}
\end{equation}
which represents a quenched pinning force acting on the interface. 
As a linear combination of a Gaussian field, the random pinning force $F_p$ is again Gaussian.
Its average is zero and its correlations are given by
\begin{equation}
 \overline{F_p(u_1,y_1) F_p(u_2,y_2)}
 =\epsilon^2\delta(y_1-y_2)\Gamma(u_2-u_1),
 \label{eq:correlationsFpGL}
\end{equation}
where the correlator along the $x$ direction is defined as
\begin{equation}
\begin{aligned}
\Gamma(u)= \gamma^2 \int_{-\infty}^{\infty} \!\! \de x \, \big({\varphi^{*}}'{\varphi^{*}}''\big)(x)\big({\varphi^{*}}'{\varphi^{*}}''\big)\big(x-u\big).\\
\end{aligned}
\label{eq:correlator}
\end{equation}

Using the explicit shape~\eqref{eq:Soliton} of the profile $\varphi^*(x)$, 
one obtains by direct computation 
\begin{equation}
\begin{aligned}
\hspace*{-12mm} 
\Gamma(u)
=& \frac{2 \alpha^3\gamma}{3 \delta^2w^3\sinh^9\left(\frac{u}{w}\right)}\left( 115 \sinh \left(\frac{u}{w}\right) \right. \\
& +90 \sinh \left(\frac{3 u}{w}\right) + 7 \sinh\left(\frac{5 u}{w}\right)\\
&-\frac{u}{w} 336\cosh \left(\frac{u}{w}\right)-\frac{u}{w}81 \cosh \left(\frac{3 u}{w}\right)\\
&\left.-\frac{u}{w}3 \cosh \left(\frac{5 u}{w}\right)\right).\
\label{eq:N(u)}
\end{aligned}
\end{equation}
The effective disorder correlations are thus short-range with a correlation length of the order of the interface width~$w$ (see also Appendix~\ref{appendix-normalized-density}).
The Fourier transform of the correlator~\eqref{eq:correlator}, defined as $\hat{\Gamma}(q)=\int_{-\infty}^{\infty} \de u \, e^{-iqu}\Gamma(u)$, is given by $\hat{\Gamma}(q)=Dg^2(q,w)$, where $D=\frac{2 \alpha^3\gamma}{9 \delta^2}$ and
\begin{equation}
 g(q,w)=\frac{\pi}{ 8 w}  (wq)^2 \left(w^2q^2+4\right) \sinh^{-1}\!\left(\frac{\pi 
   wq }{2}\right).
\end{equation}

A pinning force with correlations given by Eq.~\eqref{eq:correlationsFpGL}, for fixed $y$ and continuous $u$, may be generated by computing $F_p(u,y)=\epsilon\sqrt{\frac{D}{L_T}}\sum^{M-1}_{n=0}e^{iq_nu}g(q_n,w)z_n$, where  $q_n=\frac{2\pi}{L_{x}}n$ and $z_n$ are complex Hermitian random numbers taken from a Gaussian distributions with zero mean and unit variance, with $z_0$=0. Here, $L_x=M\delta l$ is the transverse length of the system. In Fig.~\ref{fig:Fp} we show the computed correlations of pinning forces generated with this method, for $M=10^4$, $\delta l=0.1$, $D=1$, $\epsilon=1$~\footnote{In spite of the Fourier representation,  the computational cost of obtaining the quenched random force  is very high since a Fourier transform needs to be done for each simulation cell at each simulation step (for every updated value of $u$). In order to reduce this computational cost, we  generated  4096 independent pinning forces $F_p(u,y)$ beforehand, where the $u$'s are sampled on a grid with $M=10^4$ and $\delta l=0.1$.
With these, we define an approximate $F_p(u,y)$, updated for each value of $u$ by linear interpolation during the simulation. We checked numerically that the disorder generated in this way possesses the expected correlator with satisfactory precision. For a system of length $L=256$ we checked that this method produces the same results within numerical errors compared to the full computation of the Fourier transform.}.

\begin{figure}
\begin{center}
 {\includegraphics[width=0.5\textwidth]{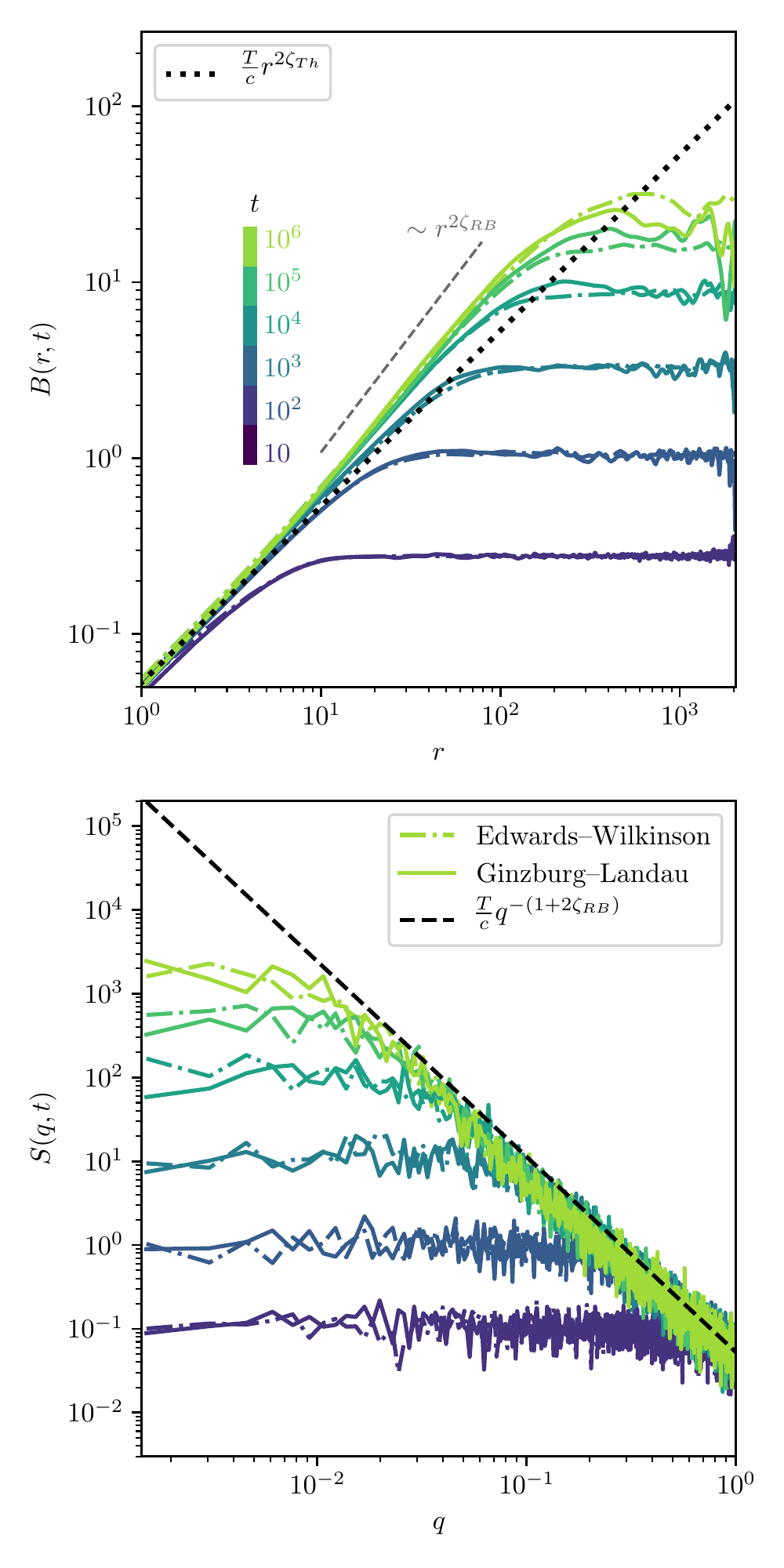}}
\end{center}
\caption{Comparison of observables for a 2D Ginzburg--Landau system (continuous lines) and for an equivalent 1D Edwards--Wilkinson system (dot-dashed lines), obtained after averaging over 10 realizations of simulations which evolved during different times $t$, indicated by different colors, at temperature $T=0.05$ and with disorder intensity $\epsilon=0.1$ starting from a completely flat configuration. On the top figure, $B(r,t)$ for the larger simulation times show deviations from the thermal regime (dotted black line). For these larger times, $B(r,t)$ is characterized by the roughness exponent $\zeta_{RB}=2/3$. On the bottom figure, we show the structure factor $S(q,t)$, defined in Eq.~\eqref{eq:S(q)}.}
  \label{fig:disorder}
\end{figure}

In Fig.~\ref{fig:disorder}, we show the excellent agreement between simulations on the 2D-GL model and on the 1D-EW model where disorder was implemented through the aforementioned method. 
At large time and large scale, the roughness function departs from the thermal behavior $\sim r^{2\zeta_{th}}$ by developing a power-law regime which is compatible with the expected scaling $\sim r^{\zeta_{RB}}$ of the so-called `random-bond' regime ($\zeta_{RB}=2/3$).
This indicates that our test of the model reduction validates a regime where disorder is relevant. 

Having established a connection between the 2D-GL and the 1D-EW models (which, however may be extended to higher dimensions, as briefly discussed at the end of Appendix~\ref{sec:GL-DES-Hamiltonian-solitonic-ansatz}) has several advantages. Exploiting the fact that an interface in a GL model behaves as one in the EW model under the elastic approximation (and small values of $T/T^\star$), allows one to avoid recomputing dynamic and static exponents of interest for the more ``realistic'' GL case. More importantly, how different quantities deviate from the expected value when the elastic limit is not satisfied may be studied in detail.

Besides, the mapping between the 2D-GL and the 1D-EW, when the elastic limit is satisfied, allows one to reduce the system size from $L_x\times L_y$ to $L_y$, and hence the computational cost~\footnote{Compared to a 2D Ginzburg--Landau simulation (solving numerically Eq.~\eqref{eq:Langevin}, with $V_\zeta$ given by Eq.~\eqref{eq:Vdisorder}), the 1D Edwards--Wilkinson Eq.~\eqref{eq:effu_disorder} is roughly 
2 times faster to solve for an equivalent evolution time $t$ with the interpolation method for the disorder implementation. We provide this information only for practical and illustrative purposes, and we do not intend to discuss acceleration techniques in the present work. The 2$\times$ speed-up of computation is a lower bound, and a baseline for further improvement of the codes.}.
In addition, from a general point of view, the model reduction allowed us to determine explicitly the disorder distribution to which the GL interface is effectively subjected to, as a result of the bulk disorder. We focused on the random bond case, but other cases, such as random-field or random-periodic disorders~\cite{giamarchi_2006_arXiv:0503437} can be treated in a similar fashion as we did.
Generically, the method we propose allows one in principle to determine the effective disorder of the EW model starting from an arbitrary disorder distribution at the bulk GL level.

\section{Conclusion and perspectives}
\label{sec:Conclusions}

Solving interface statics and dynamics beyond the elastic approximation is still a largely open theoretical and analytical problem. The disordered elastic systems theoretical framework has been proven helpful to analyze interface properties under the elastic approximation, but it can not take into account many features of experimental interfaces. A more complete description, at a large computational cost, is to use directly the Ginzburg--Landau (GL) description of the whole system (\textit{e.g.}~in 2D), 
where, by opposition to the 1D elastic line model, no assumptions need to be done over the function describing the position of the interface.

Connecting quantitatively these two descriptions has however proved elusive for extended interfaces, especially in presence of quenched disorder. We demonstrate in the present paper an analytical method to connect quantitatively the GL and the EW models with very simple assumptions. 
Compared to historical approaches that are either complex~\cite{kawasaki_dynamics_1977,kawasaki_kink_1982,10.1143PTP.67.147_kawasaki,diehl_interface_1980,PhysRevA.43.1727_grossmann,Zia1985_normalcoord,PhysRevE.64.021604_kosterlitz}
or deal with rigid walls~\cite{PhysRevLett.47.1837_bausch,PhysRevB.28.5496_grant}, 
or are more phenomenological~\cite{PhysRevB.28.2693_safran},
the method we propose has the advantage of simplicity while retaining the main features of the bulk dynamics.
We test this method by performing simulations at both levels in 2D and 1D respectively, showing how an interface in the GL model behaves. We obtain an excellent agreement with an effective elastic line in the EW model with the adequate elastic coefficients, friction and disorder distribution.

In particular, we examine the evolution in time and space of an evolving interface which is initially flat in both models by computing its spatial correlations, the so-called roughness $B(r,t)$, as a function of the evolving time of interfaces. For clean systems, we compute analytically how the roughness $B(r,t)$ of interfaces should behave under the elastic approximation, and we show how the simulated interfaces follow accurately our analytical predictions. We also probe the limit of the model reduction (which is expected to be valid in the low-temperature limit) by showing that the dynamics of the GL interface departs from the EW one at high enough temperature.
We also determined the characteristic temperature $T^\star$ below which the effective 1D description is expected to be valid.

Our method, which has been demonstrated on the time-dependent motion of a 1D interface, is quite general and can be applied to other systems. The possibility to go from the GL to the much simpler interface has a twofold interest:
\textit{(i)}~for systems for which the elastic limit is valid, it provides a path to speed up considerably the simulations compared to the higher dimensional GL description, while retaining the semi-microscopic knowledge of the parameters of the system that are more readily accessible from experiments for the GL description than for the more phenomenological interface one;
\textit{(ii)}~for systems for which the elastic limit is violated due to too large thermal noise or disorder strengths, it provides a path to quantitatively compare the direct GL simulation including all these effects with the simplified elastic description. This should help in asserting the role of ``defects'' such as overhangs, bubbles or for periodic systems with topological defects. Our approach also gives a framework to test and develop new observables to study the geometry of interfaces with overhangs and bubbles. It also serves as a tool to test how the roughness of interfaces is affected by defects.

These exciting directions go clearly beyond the reach of the present paper and will be left for future studies.

\section{Acknowledgments}
This work was supported in part by the Swiss National Science Foundation under Division II. 
N.C.~acknowledges support from the Federal Commission for Scholarships for Foreign Students for the Swiss Government Excellence Scholarship (ESKAS No.~2018.0636) for the academic year 2018-19.
V.L.~thanks the Université de Genève (where part of this work was performed) for its warm hospitality, and acknowledges support by the ERC Starting Grant No.~680275 MALIG, the ANR-18-CE30-0028-01 Grant LABS and the ANR-15-CE40-0020-03 Grant LSD.
E.A.~acknowledges support from the Swiss National Science Foundation by the SNSF Ambizione Grant PZ00P2{\_}173962.
We would also like to thank S.~Bustingorry, J.-P.~Eckmann, E.E.~Ferrero, and A.B.~Kolton for fruitful discussions related to this work. We also thank J.-P.~Eckmann for a constructive criticism of the manuscript.
The simulations were performed at the Université de Genève on the \textit{Mafalda} cluster. 

\bigskip
    
\appendix

\section{Low temperature}
\label{sec:Tstar}

In this Appendix, we determine the condition on the temperature $T$ which ensures that the thermal fluctuations of the bulk order parameter $\varphi(\mathbf r,t)$ around one of the values $\pm \varphi_0$ remain small compared to the difference of order parameter $2\varphi_0$ between the two phases.
To do so, one can write $\varphi(\mathbf r,t)=(1+\hat\varphi(\mathbf r,t))\,\varphi_0$ and determine in which regime of temperature $\hat\varphi(\mathbf r,t)$ remains much smaller than $1$ far away from the domain wall position.
Expanding the Langevin equation~\eqref{eq:Langevin} (in the absence of external field $h$), one finds
\begin{equation}
\eta \varphi_0\, \partial_t  \hat \varphi
= 
\gamma\varphi_0\nabla_{{\mathbf{r}}}^2\hat\varphi 
-2\alpha\varphi_0\,\hat \varphi + (2\eta T)^{\frac 12} \hat \xi \, , \\ 
\label{eq:Langevindeltaphihat}
\end{equation}
where the rescaled white noise $\hat\xi(\mathbf r,t)$ has correlations
$
\langle \hat \xi({\mathbf{r}},t) \hat\xi({\mathbf{r}'},t') \rangle
=
\delta^n({\mathbf{r}}'-{\mathbf{r}})\delta(t'-t)
$.
Going to Fourier space for the spatial coordinates, we see that for each mode $\mathbf q$, the Fourier transform $\hat \varphi_{\mathbf q}$ verifies an Ornstein--Uhlenbeck~\cite{PhysRev.36.823_OU} equation of the form
\begin{equation}
\partial_t  \hat \varphi_{\mathbf q}
= 
-
\frac 1 \eta
\big[
2\alpha + \gamma \mathbf q^2
\big]
\hat\varphi_{\mathbf q}
+
\Big[
\frac{2T}{\eta \varphi_0^2}
\Big]^{\frac 12} \hat \xi_{\mathbf q} \,, \\ 
\label{eq:LangevindeltaphihatFourier}
\end{equation}
with
$
\langle \hat \xi_{\mathbf{q}}(t) \hat\xi_{\mathbf{q}'}(t') \rangle
=
\delta^n({\mathbf{q}}'+{\mathbf{q}})\delta(t'-t)
$.
Its equal-time correlation function at large times is known~\cite{PhysRev.36.823_OU} and reads
\begin{equation}
    \langle\hat\varphi_{\mathbf q}(t) \hat\varphi_{\mathbf q'} (t)\rangle
    =
    \frac{T\,\delta^n({\mathbf{q}}'+{\mathbf{q}})}{(2\alpha + \gamma \mathbf q^2) \varphi_0^2}
\qquad
\text{for }
t\to\infty\,.
\end{equation}
(One finds the same result by using the Boltzmann weight and a Hamiltonian expanded quadratically close to $\varphi_0$).

Coming back to real space, for our case of interest $n=2$, \emph{i.e.}~$\mathbf r=(x,y)$, we see that, in the steady state, the equal-time correlations are logarithmically divergent (with the distance) if evaluated at two closeby points: for $t\to\infty$ and $\delta\mathbf r\to 0$, one has
\begin{equation}
    \langle
    \hat\varphi({\mathbf r},t) \hat\varphi({\mathbf r}+\delta\mathbf r,t)
    \rangle
    =
    \frac{T}{\gamma\varphi_0^2}_{\vphantom{|_{|_|}}}
    \Big(
    \text{Constant}
    +
    \log\frac{\delta\mathbf r}{w}\Big)
.
\end{equation}
In order to still get a typical temperature scale, one can take a vector $\delta\mathbf r$ of norm of the order $w=\sqrt{2\gamma/\alpha}$ and one finds
\begin{equation}
    \langle
    \hat\varphi({\mathbf r},t) \hat\varphi({\mathbf r}+\delta\mathbf r,t)
    \rangle
        \propto
    \frac{T}{\gamma\varphi_0^2}
    _{\phantom{|_{|_|}}}
\qquad
\text{for }
t\to\infty\,,
\end{equation}
up to a numerical prefactor.
Using the expression of $\varphi_0$, we thus define a characteristic temperature 
\begin{equation}
 \label{eq:defTstar}    
 T^\star = {\frac{\alpha\gamma}{\delta}}
\end{equation}
such that for $T\ll T^\star$, the typical amplitude of the thermal fluctuations of $\hat \varphi$ are small.
Note that, up to a numerical factor, one has $T^\star= w^2\, \Delta V$ with $w=\sqrt{2\gamma/\alpha}$ the lengthscale of elasticity (which also gives the domain-wall width) and $\Delta V=V(0)-V(\varphi_0)=\alpha^2/(4\delta)$ the barrier of the $\varphi^4$ potential. From the expression of the Hamiltonian, we thus see that $T^\star$ is an energy, as expected.

We also refer the reader to Ref.~\cite{bausch_effects_1991} for a study of the influence of bulk thermal fluctuations on the motion of interfaces.

\section{Solitonic ansatz in the Hamiltonian}
\label{sec:GL-DES-Hamiltonian-solitonic-ansatz}

Here we show how the connection between the GL and the DES descriptions can actually be performed directly at the level of the Hamiltonian as well, if the system is assumed to be at equilibrium.

We recall that, for the boundary conditions that we consider ${\varphi(x \pm \infty,y)=\mp \varphi_0}$,
the solitonic profile ${\varphi^*(x)}$ is the exact optimal profile at zero temperature, without disorder and in absence of external field (${T=0}$, ${\zeta \equiv 0}$, ${h=0}$).
It satisfies the extremalization condition ${\delta \mathcal{H}_{\text{GL}}[\varphi,\zeta] /\delta \varphi({\mathbf{r}}) \vert_{\varphi^*}=0}$, which translates for the Hamiltonian~\eqref{eq:Hamiltonian} into the equation ${\gamma \nabla^2 \varphi^*({\mathbf{r}}) = V_{\zeta=0}'(\varphi^*({\mathbf{r}}))}$.
As we did in Sec.~\ref{sec:FromBulktoInterface}.
we consider from now on the 2D solitonic ansatz
${\varphi_u(x,y)=\varphi^*(x-u(y))}$,
where ${\gamma \varphi{^*}''(x)=V_{\zeta=0}'(\varphi^*(x))}$ and ${\varphi^*(x \pm \infty) = \mp \varphi_0}$,
and our aim is to compute explicitly the corresponding Hamiltonian.
Since our derivation is not specific to the double-well potential ${V_{\zeta=0}(\varphi)}$, we will keep ${\varphi^*(x)}$ generic but remembering whenever needed its explicit form from Eqs.~\eqref{eq:Soliton}-\eqref{eq:SolitonParameters},
${\varphi^*(x-u) = - \varphi_0 \tanh ((x-u)/w)}$
with ${\varphi_0 = \sqrt{\alpha/\delta}}$ and ${w = \sqrt{2 \gamma / \alpha}}$.
We will moreover need the following definitions of constants, slight generalisations of Eqs.~\eqref{eq:defN1-DW}-\eqref{eq:defN2N3-DW}:
\begin{equation}
\begin{split}
 \mathcal{N}_1(u)
 	& \equiv \int \de x \, \argc{ \varphi{^*}'(x-u)}^2
 	\stackrel{(x \in \mathbb{R})}{=} \mathcal{N}_1
 	\, ,
 \\
 \mathcal{N}_2(u)
 	& \equiv \int \de x \, \varphi{^*}'(x-u) \, \varphi{^*}''(x-u)
 	\\
 	&= \int \de x \, \partial_x \argc{ \frac12 \varphi{^*}'(x-u)^2}
 	\stackrel{(x \in \mathbb{R})}{=} \mathcal{N}_2
 	\, ,
 \\
 \mathcal{N}_3(u)
 	& \equiv \int \de x \, \varphi{^*}'(x-u)
 	\stackrel{(x \in \mathbb{R})}{=} \mathcal{N}_3
 	\, ,
\end{split}
\label{eq-def-mathcalN123}
\end{equation}
with ${\mathcal{N}_2=0}$ since ${\varphi{^*}'(x \to \pm \infty)=0}$, and specifically for the double-well potential ${\mathcal{N}_1=\frac43 \varphi_0^2/w}$ and ${\mathcal{N}_3=-2\varphi_0}$ (for the boundary conditions ${\varphi^*(x \to \pm \infty)=\mp \varphi_0}$).
We emphasize that we are able to get rid of the dependence on $u$ in Eq.~\eqref{eq-def-mathcalN123} if ${\varphi{^*}'(x)}$ decays sufficiently fast with respect to system size in the $x$-direction; this becomes exact for ${x \in \mathbb{R}}$, but should be kept in mind otherwise.

We compute explicitly the energy associated to the ansatz ${\varphi_u(x,y)}$:
\begin{equation}
\begin{split}
 & \mathcal{H}_{\text{GL}} \argc{\varphi_u,\zeta}
 = \int \de y \, \de y \arga{
 		\frac{\gamma}{2} \argc{\nabla \varphi_u (x,y)}^2
 		+ V_{\zeta} \argp{\varphi_u (x,y)}
 		}
 \\ 
 &= \int \de y \, \de x \, \left\lbrace
			\frac{\gamma}{2} \argc{ \argp{\partial_x \varphi_u(x,y)}^2 + \argp{\partial_y \varphi_u(x,y)}^2} \right.
 \\
	& \left. \phantom{\frac{\gamma}{2}} \quad\quad\quad\quad\quad\quad + \argp{1+\epsilon \zeta (x,y)} V_{\zeta=0}\argp{\varphi_u(x,y)} \right\rbrace
 \\
 &= \int \de y \, \left\lbrace
 		\frac{\gamma}{2} \mathcal{N}_1\argp{u(y)} \argc{\argp{\partial_y u(y)}^2 + 1}
  	\right.
 \\
	& \left. \phantom{\frac{\gamma}{2}} \quad\quad\quad\quad\quad\quad + \argp{1+\epsilon \zeta (x,y)} V_{\zeta=0}\argp{\varphi^*\argp{x-u(y)}} \right\rbrace
 \\
 & \equiv \int \de y \, \argc{
 		\frac{c}{2} \argp{\partial_y u(y)}^2 + U_p \argp{u(y),y}} + \mathcal{C}
 \\
 & \equiv \mathcal{H}_{\text{DES}} \argc{u,U_p} + \mathcal{C}
 \, .
\end{split}
\label{eq-explicit-GL-Hamiltonian-solitonic-ansatz}
\end{equation}
In the last two steps, we have identified the DES elastic constant and the effective pinning potential, respectively:
\begin{equation}
\begin{split}
 & c \equiv \gamma \mathcal{N}_1 \, ,
 \\
 & U_p(u,y) \equiv \epsilon \int \de x \, \zeta(x,y) \, V_{\zeta=0} \argp{\varphi^*\argp{x-u}} \, ,
\end{split}
\end{equation}
and an additive term independent of $u$ thanks to ${x \in \mathbb{R}}$:
\begin{equation}
\begin{split}
 \mathcal{C}
 & \equiv \int \de y \, \argc{
	\frac{\gamma}{2} \mathcal{N}_1 \argp{u(y)} + \int \de x \, V_{\zeta=0}\argp{\varphi^*(x-u(y))}
 }
 \\
 &= \int \de y \, \argc{
	\frac{\gamma}{2} \mathcal{N}_1 + \int \de x \, V_{\zeta=0}\argp{\varphi^*(x)}
	}
 \, .
\end{split}
\end{equation}
Although for an infinite system size ${\mathcal{C}}$ might diverge, it is a well-defined finite constant for any finite system size, and as such it can be safely removed by normalization of the Gibbs--Boltzmann weight from the definition of the actual DES Hamiltonian ${\mathcal{H}_{\text{DES}} \argc{u,U_p}}$. 
Physically $\mathcal{C}$ corresponds to the elastic energy associated to the gradient in the $x$ direction (${\propto (\partial_x \varphi_u(x,y))^2}$) and the energy associated to the bare double-well potential ${V_{\zeta=0}}$ (since the two phases ${\pm \varphi_0}$ are of equal energy and the domain wall is spatially symmetric in $x$);
if we assume the same solitonic profile ${\forall y}$, as we have done with the ansatz ${\varphi_u(x,y)}$, then these two contributions to the energy do not depend on $u$ and thus are indeed irrelevant in an effective DES description of the system.

The pinning potential ${U_p(y,u)}$ is linear in the underlying GL disorder ${\zeta}$, consequently it inherits its Gaussian distribution,
with zero mean ${\overline{U_p \argp{u,y}}= 0}$
and two-point correlation:
\begin{equation}
\begin{split}
 & \overline{U_p \argp{u,y} U_p \argp{u',y'}}
	\equiv R_w (u,u') \, \delta( y-y')
 \, ,
 \\
 & R_w (u,u')
 	\equiv \epsilon^2 \argc{\int \de x \, V_{\zeta=0} \argp{\varphi^*(x-u)}
 V_{\zeta=0} \argp{\varphi^*(x-u')}}
 \,
 \\
	& \quad\quad\quad\quad\, = \frac{\epsilon^2\gamma^2}{4} \int \de x \, \varphi{^*}'(x-u)^2 \varphi{^*}'(x-u')^2
 \, .
\end{split}
\end{equation}
We used in the last equality the defining relation ${\gamma \varphi{^*}''=V_{\zeta=0}'(\varphi^*)}$ (but no need to specify ${V_{\zeta}(\varphi)}$), and this allows us to notice that ${R_w(u,u')=R_w(u-u')}$.
In order to reconnect with the pinning force ${F_p(u,y)}$ defined in Eq.~\eqref{eq:Fp}, note that
\begin{equation}
\begin{split}
 & F_p(u,y) = - \partial_u U_p(u,y)
 \\
 &= \epsilon \int \de x \,  \zeta (x,y)  \, V_{\zeta=0}'\argp{ \varphi{^*}\argp{x-u(y)}} \varphi{^*}'\argp{x-u}
 \\
 & = \epsilon \gamma \int \de x \, \zeta(x,y) \, \varphi{^*}''\argp{x-u} \varphi{^*}'\argp{x-u}
 \\
 & \stackrel{(x \in \mathbb{R})}{=} \epsilon \gamma \int \de x \, \zeta(x+u,y) \, \varphi{^*}''\argp{x} \varphi{^*}'\argp{x}
 \, .
\end{split}
\end{equation}
And as for the force-force correlator~\eqref{eq:correlationsFpGL}, we have similarly:
\begin{equation}
\begin{split}
 \overline{F_p \argp{u,y} F_p \argp{u',y'}}
 & = \partial_u \partial_{u'} \overline{U_p(u,y) U_p(u',y')}
 \\
 &= - R_w''(u-u') \, \delta (y-y')
 \\
 &\stackrel{\eqref{eq:correlationsFpGL}}{\equiv} \epsilon^2 \, \Gamma(u-u') \, \delta (y-y')
 \, ,
\end{split}
\end{equation}
with the correlator ${\Gamma(u-u')}$ introduced and discussed in Sec.~\ref{sec:Disorder}.

The bottom line of Eq.~\eqref{eq-explicit-GL-Hamiltonian-solitonic-ansatz} is that, with the solitonic ansatz ${\varphi_u(x,y)}$, the GL Hamiltonian reduces exactly (without any approximation) into a DES Hamiltonian function of $u(y)$, of elastic constant $c$ and pinning potential $U_p$ (with the two-point correlator ${R_w(x)}$):
\begin{equation}
 \mathcal{H}_{\text{GL}} \argc{\varphi_u,\zeta}\Big\vert_{\alpha,\gamma,\delta,\epsilon}
 \equiv \mathcal{H}_{\text{DES}} \argc{u,U_p}\Big\vert_{c,R_w(x)}  + \mathcal{C}
 \, .
\label{eq-nutshell-Hamiltonian-ansatz}
\end{equation}
This also implies that, if we need to determine the DES force acting on the displacement field ${u(y)}$ in its associated Langevin dynamics, we must use the functional `chain rule' as follows:
\begin{equation}
\begin{split}
 & F_p \argp{u(y),y}
 = - \frac{\delta \mathcal{H}_{\text{GL}} \argc{\varphi_u,\zeta}}{\delta u(y)}
 \\
 &= \int \! \de x \int \! \de y' \, \argc{- \frac{\delta \mathcal{H}_{\text{GL}} \argc{\varphi_u,\zeta}}{\delta \varphi_u (x,y')}} \, \frac{\delta \varphi_u(x,y)}{\delta u(y')} \, \underbrace{\frac{\delta u(y')}{\delta u(y)}}_{\delta (y-y')}
 \\
 &= \int \de x \, \argc{- \frac{\delta \mathcal{H}_{\text{GL}} \argc{\varphi_u,\zeta}}{\delta \varphi_u (x,y)}} \, \argc{- \varphi{^*}'(x-u(y))}
 \, ,
\end{split}
\label{eq-chain-rule-functional-DEs-force}
\end{equation}
which firmly supports our procedure to go from Eq.~\eqref{eq:Langevinansatz1bis} to Eq.~\eqref{eq:EW}, namely to multiply by the profile density ${\varphi{^*}'(x-u(y))}$ and perform the integration ${\int \de x \, (\dots)}$.

In addition, our physical motivation for even considering ${\mathcal{H}_{\text{GL}} \argc{\varphi_u,\zeta}}$ is that, at sufficiently low temperature, the statistical average over thermal fluctuations should be dominated by the optimal profile.
In a nutshell, this assumption can be formalized as follows (${\mathcal{O}}$ being an observable without an explicit dependence on the disorder):
\begin{equation}
\begin{split}
 &\overline{\moy{\mathcal{O}}}
 = \overline{\int \mathcal{D} \varphi \, \mathcal{P} [\varphi ,\zeta] \, \mathcal{O} [\varphi]}
 \\
 & \quad\quad \stackrel{[\text{ansatz } \varphi_u(x,y) ]}{\approx}
 		\overline{\int \mathcal{D} u \, \mathcal{P} [u, \varphi^* ,\zeta]  \, \mathcal{O} [u,\varphi^*]}
\end{split}
\end{equation}
with
${\mathcal{P} [\varphi ,\zeta] \propto \exp \arga{-\frac{1}{T} \mathcal{H}_{\text{GL}}[\varphi,\zeta]}}$
being the Gibbs--Boltzmann weight and thanks to our result~\eqref{eq-nutshell-Hamiltonian-ansatz}
\begin{equation}
\begin{split}
\mathcal{P} [u, \varphi^* ,\zeta]
 	&\propto \exp \arga{-\frac{1}{T} \mathcal{H}_{\text{GL}}[u, \varphi^*,\zeta] } \Big\vert_{\alpha,\gamma,\delta, \epsilon}
 	\\
 	& \propto \exp \arga{-\frac{1}{T} \mathcal{H}_{\text{DES}}[u, U_p]} \Big\vert_{c,R_w(x)}
 	\, .
\end{split}
\end{equation}
This model reduction of the equilibrium path integral, using only the solitonic ansatz ${\varphi_u(x,y)}$, should be modified for slightly higher temperature by taking into account at first the thermal fluctuations both in $u$ around the ${\varphi_u}$ and of the profile ${\varphi^*}$ itself.

Finally, note that the derivation presented in this appendix can straightforwardly be generalized to higher dimensions ${{\mathbf{r}} = (x,{\mathbf{y}}) \in \mathbb{R}^d}$, for an interface parametrized by a displacement field ${u({\mathbf{y}}) \in \mathbb{R}}$ along the direction ${\hat{{\mathbf{x}}}}$ with ${{\mathbf{y}}\in \mathbb{R}^{d-1}}$ the `internal' coordinate in the plane ${\perp \hat{{\mathbf{x}}}}$.
Using the solitonic ansatz ${\varphi_u(x,{\mathbf{y}})=\varphi^*(x-u({\mathbf{y}}))}$, we obtain:
\begin{equation}
\begin{split}
 \mathcal{H}_{\text{GL}} \argc{\varphi_u,\zeta}
    =& \! \int \!\de {\mathbf{y}} \, \argc{
 		\frac{c}{2} \argp{\nabla_{{\mathbf{y}}} u({\mathbf{y}})}^2 + U_p \argp{u({\mathbf{y}}),{\mathbf{y}}}} + \mathcal{C}
 \\
 \equiv & \, \mathcal{H}_{\text{DES}} \argc{u,U_p} + \mathcal{C}
 \, ,
 \\
 - \frac{\delta  \mathcal{H}_{\text{GL}} \argc{\varphi_u,\zeta}}{\delta u({\mathbf{y}})}
    & = c \nabla^2_{{\mathbf{y}}} u ({\mathbf{y}}) 
        + F_p \argp{u({\mathbf{y}}),\mathbf{y}}
 \\
  & = c \nabla^2_{{\mathbf{y}}} u ({\mathbf{y}}) 
        - \partial_u U_p \argp{u,\mathbf{y}} \vert_{u=u({\mathbf{y}})}
 \, ,
\end{split}
\label{eq-explicit-GL-Hamiltonian-solitonic-ansatz-high-dim}
\end{equation}
where the only modification in the two-point correlators for ${u_p}$ and ${F_p}$ consists in replacing the 1D ${\delta (y-y')}$ by its multi-dimensional counterpart ${\delta^{d-1} ({\mathbf{y}} -{\mathbf{y}}')}$.

\section{Path-integral approach}
\label{sec:PI-nodis}

For the Ginzburg--Landau Langevin dynamics~\eqref{eq:Langevin} in absence of disorder (${\zeta \equiv 0}$), the trajectorial probability on a time window $[0,t_\ff]$ writes
\begin{align}
  \mathbb P[\varphi] &\propto e^{-S[\varphi]}
\\
  S[\varphi] &= 
\frac{1}{4\eta T}
\int_0^{t_\ff} \de t \int \de x \, \de y\, \Big(
\eta \partial_t \varphi +\frac{\delta \mathcal H_{\text{GL}}[\varphi]}{\delta \varphi}
\Big)^2,
\end{align}
where the action $S[\varphi]$ is given in its Onsager--Machlup form.
Using the solitonic ansatz for
${\varphi(x,y,t)}$, ${\varphi_u(x,y,t)\equiv\varphi^*(x-u(y,t))}$, the action $S[\varphi_u]$ represents (through $e^{-S[\varphi_u]}$) the weight of the profile $\varphi_u$ among every other possible profile $\varphi(x,y,t)$.
Integrating over the coordinate $x$, one finds
by direct computation that
\begin{align}
\!\!
S[\varphi_u]
=
\frac{1}{4 T \eta \mathcal{N}_1}
\int_0^{t_\ff}\!\!\!\! \de t\!\!
\int \!\! \de y
\bigg\{
\!
&\bigg(
  \eta \mathcal{N}_1
  \partial_t u
- 
  \gamma \mathcal{N}_1
  \partial_y^2 u
- 
  h \mathcal{N}_3
\bigg)^2
\nonumber
\\
&
\quad\quad
+ \frac{16}{45}V_0^2 \big( \partial_y u )^4
\bigg\}\,,
\label{eq:Sofphiu}
\end{align}
where according to Eqs.~\eqref{eq:defN1-DW}-\eqref{eq:defN2N3-DW} we have
${\mathcal{N}_1 = \frac43 \varphi_0^2/w}$ and
${\gamma \mathcal{N}_1 = \frac23 V_0 w}$,
with
$V_0=\alpha^2/\delta$ the amplitude of the $\varphi^4$ potential.
The quartic term $\propto (\partial_y u)^4$ indicates that such an action is not exactly in the expected form of an action corresponding to a Langevin equation for the evolution of $u(y,t)$ with a Gaussian white noise.
A similar quartic term occurs when implementing such procedure for the noisy Landau--Lifschitz--Gilbert bulk dynamics~\cite{PhysRevLett.98.056605_duine}.
Such supplementary terms remind us that the zero-noise ansatz profile $\varphi_u$ is not the exact profile of the bulk model:
this corresponds to the fact, discussed in the main text, that at the Langevin level, Eq.~\eqref{eq:Langevinansatz1bis} is not exact.
In Eq.~\eqref{eq:Sofphiu}, for small displacements $u$, it can be neglected and the effective action for the position $u(y,t)$ of the interface reads

\begin{align}
\mathcal S^\eff[u] 
&= 
\frac{1}{4\tilde\eta T}
\int_0^{t_\ff} \de t 
\int \de y\,
\Big(
\tilde\eta \partial_t u - c \partial_y^2 u - F
\Big)^2
\:.
\label{eq:defSueff}
\end{align}
It corresponds to an Edwards--Wilkinson equation for $u(y,t)$, of the form~\eqref{eq:EW}, with the effective friction coefficient $\tilde \eta$, elasticity constant $c$ and external force $F$ as the ones we found in Eq.~\eqref{eq:ParametersAnalogies} using the direct Langevin approach.

\section{Interface normalized density}
\label{appendix-normalized-density}

The solitonic ansatz ${\varphi_u(x,y)=\varphi^*(x-u(y))}$ is at the core of our procedure for connecting the GL to the DES description, with ${\varphi^*}$ being the exact optimal profile at zero temperature, without disorder and in absence of external field.

The derivative of this profile, ${\varphi{^*}'(x)}$, can be interpreted straightforwardly as an unnormalized `density' of the domain wall (or interface) between the two phases ${\pm \varphi_0}$ imposed by the boundary conditions (as illustrated in the inset of Fig.~\ref{fig:wall3d}).
Its normalized counterpart is then defined as:
\begin{equation}
\begin{split}
 \rho_{w}(x)
 	&\equiv \frac{\varphi{^*}'(x)}{\int_{\mathbb{R}} \de x' \, \varphi{^*}'(x')}
 	\stackrel{\eqref{eq-def-mathcalN123}}{=} \frac{\varphi{^*}'(x)}{\mathcal{N}_3}
  	= \frac{1}{w} \rho_1 (x/w)
 \\ 	
 	&\stackrel{\eqref{eq:Soliton}}{=}
 		\underbrace{\frac{\varphi_0}{w \cosh^2 (x/w)}}_{=\varphi{^*}'(x)} \frac{1}{2\varphi_0}
 	= \frac{1}{2 w \cosh^2 (x/w)}
 \, ,
\end{split}
\label{eq-normalised-density-rho}
\end{equation}
and the last line is specific to the double-well potential ${V_{\zeta=0}(\varphi)}$.
Thereafter we keep the profile $\varphi^*$ and its corresponding density generic, but keeping in mind that
\textit{(i)}~$\varphi^*$ is an odd function (inherited for instance from the \textit{symmetric} double well),
and \textit{(ii)}~it satisfies ${\gamma \partial_x^2 \varphi^*=V'_{\zeta=0}(\varphi^*)}$.
Note that we denoted in Ref.~\cite{agoritsas2010} this normalized density by ${\rho_{\xi}(x)}$ where $\xi$ corresponds to the effective `width' and thus can be identified (up to an arbitrary numerical constant) with the parameter ${w}$.

We can consequently rewrite, for ${\lbrace \mathcal{N}_1,\mathcal{N}_2 \rbrace}$ whose definitions are recalled in Eq.~\eqref{eq-def-mathcalN123}, and with ${\valabs{\mathcal{N}_3}=2\varphi_0}$:
\begin{equation}
\begin{split}
 \mathcal{N}_1
 	=& \mathcal{N}_3^2 \int_{\mathbb{R}} \de x \, \argc{ \rho_w(x)}^2
 	= \frac{\mathcal{N}_3^2}{w} \int_{\mathbb{R}} \de \tilde{x} \, \argc{ \rho_1(\tilde{x})}^2
 \\
 \mathcal{N}_2
 	=& \mathcal{N}_3^2 \int \de x \, \partial_x \argc{ \frac12 \rho_w(x)^2}
	\\ 	
 	& \quad\quad\quad = \frac{\mathcal{N}_3^2}{w^2} \int_{\mathbb{R}} \de \tilde{x} \, \partial_{\tilde{x}} \argc{ \frac12 \rho_1(\tilde{x})^2}
 	= 0
 \, ,
\end{split}
\end{equation}
so the specific functional form of ${\varphi{^*}'}$ will fix the numerical factor ${\int_{\mathbb{R}} \de \tilde{x} \, \argc{ \rho_1(\tilde{x})}^2}$ but the overall dependence ${\mathcal{N}_1 \propto \mathcal{N}_3^2/w}$ will not change.

Using the definition of ${\rho_w(x)}$, the disorder correlator ${R_x(u,u')}$ can similarly be rewritten as
\begin{equation}
\begin{split}
 R_w (u,u')
 &= \frac{\epsilon^2\gamma^2}{4} \mathcal{N}_3^4 \int \de x \, \rho_w(x-u)^2 \rho_w(x-u')^2
 \\
 &= \frac{\epsilon^2\gamma^2}{4} \frac{\mathcal{N}_3^4}{w^3} \int \de \tilde{x} \, \rho_1\argp{\tilde{x}-u/w}^2 \rho_1 \argp{\tilde{x}-u'/w}^2
 \\ 
 &= \frac{1}{w} R_1 \argp{\frac{u}{w},\frac{u'}{w}} \, ,
\end{split}
\end{equation}
and we can further define the strength of disorder and the normalized correlator, respectively, as:
\begin{equation}
\begin{split}
 & D^{\text{norm}} \equiv \int_\mathbb{R} \de x \, R_w(x)
 \, ,
 \\
 & R^{\text{norm}}_w(u,u') \equiv R_w(u,u')/D
\, .
\end{split}
\end{equation}
The strength of disorder becomes:
\begin{equation}
\begin{split}
 D^{\text{norm}}
 &= \frac{\epsilon^2\gamma^2}{4} \mathcal{N}_3^4  \int_{\mathbb{R}^2} \de x \, \de x' \, \rho_w(x'-x)^2 \rho_w(x')^2
 \\ 
 &= \frac{\epsilon^2\gamma^2}{4} \frac{\mathcal{N}_3^4}{w^2}  \underbrace{\int_{\mathbb{R}^2} \de \tilde{x} \, \de \tilde{x}' \, \rho_1(\tilde{x}'-\tilde{x})^2 \rho_1(\tilde{x}')^2}_{= \text{numerical prefactor}}
\end{split}
\end{equation}
which yields, for the normalized density associated to the double-well potential ${\rho_1(\tilde{x})=\frac{1}{2 \cosh^2(\tilde{x})}}$:
\begin{equation}
\begin{split}
 & D^{\text{norm}}
 	= \frac{\epsilon^2\gamma^2}{4} \frac{\mathcal{N}_3^4}{w^2}  \frac19
 	= \frac{4 \epsilon^2\gamma^2}{9} \frac{\varphi_0^4}{w^2}
 	= \epsilon^2 \, \frac29 \frac{\alpha^3 \gamma}{\delta^2}
 \, ,
 \\
 & R^{\text{norm}}_1 \argp{\tilde{u}}
 	=  9  \int_{\mathbb{R}} \de \tilde{x} \, \frac{1}{2 \cosh^2 \argp{\tilde{x}-\tilde{u}}} \frac{1}{2 \cosh^2 \argp{\tilde{x}}}
 \, ,
 \\
 & c \equiv \gamma \mathcal{N}_1
 	= \frac{2 \sqrt{2}}{3 \delta} \sqrt{\alpha^3 \gamma}
 \, .
\end{split}
\end{equation}
All this construction can nevertheless be generalized to other density functionals for ${\rho_w(x)}$, modifying only the different numerical prefactors: although qualitatively irrelevant, those are crucial for achieving a quantitative agreement, as the one we have been seeking in this work.



%

\end{document}